\documentclass[aip,amsmath,amssymb,reprint]{revtex4-1}
\usepackage{mathtools}
\usepackage{epsfig}
\usepackage{graphicx}
\usepackage[english]{babel}
\usepackage{xcolor} 
\usepackage{soul}
\usepackage[unicode=true]{hyperref}
\usepackage[thinc]{esdiff} 
\hypersetup{breaklinks=true, colorlinks=true, urlcolor= blue, citecolor=blue}
\begin{document}

\title{Disorder induced melting and glass formation in a one-component Lennard-Jones system}

\author{Saumya Suvarna}
\affiliation{Department of Physics, Birla Institute of Technology Mesra, Ranchi  835215, India}
\author{Prabhat K. Jaiswal}
\affiliation{Department of Physics, Indian Institute of Technology, Jodhpur, Karwar 342030, India}
\author{Madhu Priya}
\email[Corresponding author:\;]{madhupriya@bitmesra.ac.in}
\affiliation{Department of Physics, Birla Institute of Technology Mesra, Ranchi  835215, India}

\date{\today}
\begin{abstract}
Identifying the conditions under which glass formation occurs is crucial for a fundamental understanding of the glass transition mechanism. Pure liquids devoid of any frustration avoid glass transition and undergo crystallization. In this work, we investigate a one-component liquid interacting via the Lennard-Jones potential in two dimensions, where disorder is introduced through pinning, a protocol in which a fixed fraction of particles is immobilized at positions selected from an equilibrium configuration. By employing molecular dynamics simulation, we systematically study the influence of pinning concentration on both structural and dynamical properties. Structural properties quantified by radial distribution function and hexatic-order parameter display a systematic decrease with a rise in pinning concentration. However, the dynamical properties such as the fragility index and the late-time mean squared displacement exhibit a non-monotonic trend as the concentration of pinned particles increases. A moderate concentration of pinned particles helps prevent crystallization and facilitates particle motion. A further rise in the number of pinned particles suppresses particle mobility, leading to a reduction in the overall dynamics of the system. These simulation results are in good agreement with experimental observations on colloidal suspensions confined between glass coverslips, where particles are immobilized. Our findings demonstrate the pivotal role of pinning in controlling the phase behavior of simple liquids and validate the unique dynamical features of two-dimensional liquids with pinned particles.
\end{abstract}

\maketitle

\section{INTRODUCTION}
The glass transition remains one of the most challenging problems in condensed-matter physics, characterized by a dramatic increase in relaxation times without an accompanying thermodynamic phase transition \cite{Debenedetti2001,Berthier2011_rmp}. Understanding the precise conditions that govern glass formation \cite{MP_2007, MP_2014, MP_2015, PKJ_2016_PRL} remains a central challenge in materials science \cite{chaikin1995principles}. The application of glasses range from optical and electronic devices \cite{2009Masai_semiconductor, ELMETWALLY2022_optoelectronic} to coatings \cite{Majumdar2001_glass_coating,Rozenenkova2016_coating} and structural materials \cite{Debenedetti2001}. In conventional glass-forming systems, frustration arising from compositional disorder \cite{ka_2009,bipolyLJ}, competing length scales, confinement \cite{Gallo2002}, or random pinning plays a crucial role in suppressing crystallization and enabling access to deeply supercooled states. Monodisperse liquids, in contrast, lack inherent sources of frustration and therefore crystallize readily \cite{suvarna2024role}, making them unsuitable for probing glassy dynamics unless an external mechanism is introduced to hinder ordering.

In recent times, pinning a fraction of constituent particles has emerged as a powerful tool to induce frustration and study glassy dynamics \cite{Cammarota2012,Karmakar_2013}. The concept of pinning, whether achieved by freezing particles at random \cite{2014-Kob} or at specified locations \cite{2012BERTHIER,Chakrabarty_2015}, has significantly advanced our understanding of slow relaxation \cite{2015Li}, dynamic heterogeneity \cite{Bhowmik2016}, and the interplay between structure and dynamics in supercooled liquids \cite{Sun_2021}. Pinning introduces quenched disorder into the system, promoting kinetic arrest by hindering long-range order \cite{Cammarota2012}, thereby providing a valuable testing ground for probing theoretical predictions of the glass transition \cite{Krakoviack_2014,Szamel_2013} and exploring the ability to control material properties.

A growing body of both numerical \cite{2012BERTHIER,Jack_2013,2013-Kob-PRL,Karmakar_2013,Hocky_2014,Fullerton_2014, 2014-Kob, pinning-2015-misaki-ozawa, 2015Li,Chakrabarty_2015,Bhowmik2016,2019BHOWMIK,Dattani2023} and experimental \cite{Gokhale_2014,Sun_2021} works have highlighted the rich phase behavior and unique dynamical regimes accessible in pinned systems. Random pinning has been extensively used to investigate point-to-set correlations \cite{2012BERTHIER}, configurational entropy \cite{pinning_entropy_SK}, and predictions of the Random First-Order Transition theory \cite{Cammarota2012}, particularly in binary mixtures where crystallization is naturally suppressed. Amorphous-wall pinning techniques have further revealed how confinement and frozen boundaries influence static length scales \cite{2002Scheidler} relaxation mechanisms \cite{Kim2003}, and dynamic properties \cite{Gallo2002,2003Pallo}. In three dimensions, Karmakar and Parisi investigated the effect of random pinning on the structure and dynamics of monodisperse particles interacting via purely repulsive potentials \cite{Karmakar_2013}. In a related effort, Zhou and Milner employed a crystal-avoiding hybrid Monte Carlo approach to extract static length scales in a monodisperse hard-sphere system with template pinning \cite{Zhou_2016_template_pin}. Despite this progress, the majority of numerical studies focus on three-dimensional or inherently disordered models, while the literature on two-dimensional monodisperse Lennard-Jones liquids with pinning remains limited.



Recent experimental study has revealed an intriguing pathway to glass formation originating from an ordered crystalline state in the presence of quenched disorder introduced via pinning \cite{Sun_2021}. In particular, the observation of a crystal–hexatic–glass transition, accompanied by a non-monotonic evolution of dynamical properties with increasing pinning density, represents a fundamentally different route to glass formation compared to the conventional liquid-to-glass transition. Despite these important experimental findings, a comprehensive numerical understanding of the underlying mechanisms remains lacking.  Motivated by this open question, the present work aims to fill this research gap by performing a detailed molecular dynamics–based computational study of a model system with quenched disorder introduced through particle pinning. By systematically varying the pinning density and analyzing both structural and dynamical observables, this study seeks to provide a microscopic explanation for the emergence of glassy dynamics from an initially ordered phase and to rationalize the reported non-monotonic dynamical behavior \cite{Sun_2021}. Through this approach, the present work not only complements existing experimental observations but also advances the theoretical framework for understanding the role of pinning in disorder-driven phase transitions and glass formation in low-dimensional systems.

The work of Christoph Dellago has been central to the development of modern computational statistical mechanics, particularly through his seminal contributions to the understanding of rare events, free-energy barriers, and molecular pathways. His development of transition path sampling \cite{Dellago-book1} has enabled detailed investigations of nucleation, phase transitions, and other activated processes \cite{Dellago-JCP1,Dellago-JCP2}. Complementary advances in structural order parameters have further deepened understanding of local ordering in crystalline and amorphous systems \cite{Dellago-crystalline-structure-jcp}. Together, these contributions have had a lasting impact on how structure, dynamics, and constraints are analyzed in complex molecular systems. In this spirit, the present study employs molecular dynamics simulations to examine how disorder introduced by pinning influences structure and dynamics in a minimal two-dimensional model system.



The remainder of this paper is organized as follows. Section II describes the model system and simulation protocol. Section III presents our results on structural and dynamic properties as a function of pinning concentration. Finally, in Section IV, we summarize our findings and discuss their implications for understanding glass formation in constrained systems.

\section{METHODOLOGY}
In this study, we perform molecular dynamics simulations in LAMMPS \cite{thompson2022lammps} software to study the structure and dynamics of a pure liquid consisting of pinned particles. We consider a system of $1000$ particles at a high particle density $\rho=N/V=0.85$ in an NVT ensemble of side length $L=34.3$ in two dimensions. It has been observed that the glass transition in a two-dimensional identical-particle system occurs around this density in both numerical \cite{2D_density} and experimental studies \cite{Sun_2021}.  The initial configuration of this identical particle system is generated using the PACKMOL\cite{Packmol} software. 
 
The interactions between particles are governed by the Lennard-Jones (LJ) potential \cite{lennard1924determination,lennard1931cohesion}, expressed as,
\begin{equation}
    U(r) = 4 \varepsilon \left[ \left( \frac{\sigma}{r} \right)^{12} - \left( \frac{\sigma}{r} \right)^6 \right]
\end{equation}

The particles are identical in size and mass, with $\sigma$ representing the diameter and $m$ denoting the mass of each particle. The strength of the interparticle interaction, $\varepsilon$, is set to unity for all pairs of particles. All physical quantities presented here are expressed in reduced Lennard-Jones (LJ) units, where $\sigma$, $\varepsilon$, $\varepsilon/k_B$ and $\sqrt{m\sigma^{2}/\varepsilon}$ serve as the units of length, energy, temperature and time, respectively. The potential is shifted and truncated at $r_{cut}=3.0\sigma$ for enhanced computational efficiency \cite{allen2017computer}.

The equations of motion are integrated using the velocity-Verlet algorithm \cite{vvint} with a time step of \( dt = 10^{-3}\tau_{LJ} \), and periodic boundary conditions are applied in both directions. The system is first equilibrated at $T = 5$ to remove any initial spatial correlations. Following this, a fraction of particles is pinned at regular intervals using the template pinning protocol \cite{Zhou_2016_template_pin}. In this study, we have considered the concentration of pinned particles $c=1\%,2\%,4\%,5\%,10\%,20\%,25\%,33\%$ and $50\%$. To compare our results with a reference system without any pinned particles, simulations for $c=0\%$ have also been performed. After pinning, the particles are then re-equilibrated at the same temperature for a long duration, and then cooled with a cooling rate of \( 10^{-3}/\tau_{LJ} \) to reach the desired temperature. During equilibration, temperature is controlled using the Berendsen thermostat \cite{berend-thermostat}. Nos\'e--Hoover thermostat \cite{Nose1,Nosé2} is employed during the production runs to maintain a constant temperature. Statistical accuracy is ensured by averaging over sufficient independent simulation runs.



\section{RESULTS}
Upon cooling, two-dimensional systems exhibit transitions from liquid to crystalline state mediated by an hexatic-phase \cite{halperin1978theory}, characterized by quasi-long-range orientational order and short-range positional correlations. The presence of quenched disorder can significantly modify these transitions, influencing both the degree of structural ordering and the nature of particle relaxation. To characterize these effects in a systematic manner, it is essential to analyze the structural and dynamical observables that capture ordering at different length and time scales. We present a detailed examination of the equilibrium structure using radial and orientational order parameters, followed by an analysis of the relaxation dynamics and transport properties.



\subsection{RADIAL DISTRIBUTION FUNCTION}
The radial distribution function (RDF), $g(r)$, quantifies the likelihood of finding a particle at a distance $r$ from a given reference particle. The RDF is calculated as \cite{hansen2013theory},
\begin{equation}
    g(r) = \frac{1}{2\pi\rho N r\Delta r} \left\langle \sum_{i=1}^{N} \sum_{j\neq i}^{N} \delta (r - r_{ij}) \right\rangle,
\end{equation}
where $r_{ij}$ denotes the distance between particles $i$ and $j$, and $\rho$ is the number density of the system.

\begin{figure}
    \centering
    \includegraphics[width=1\linewidth]{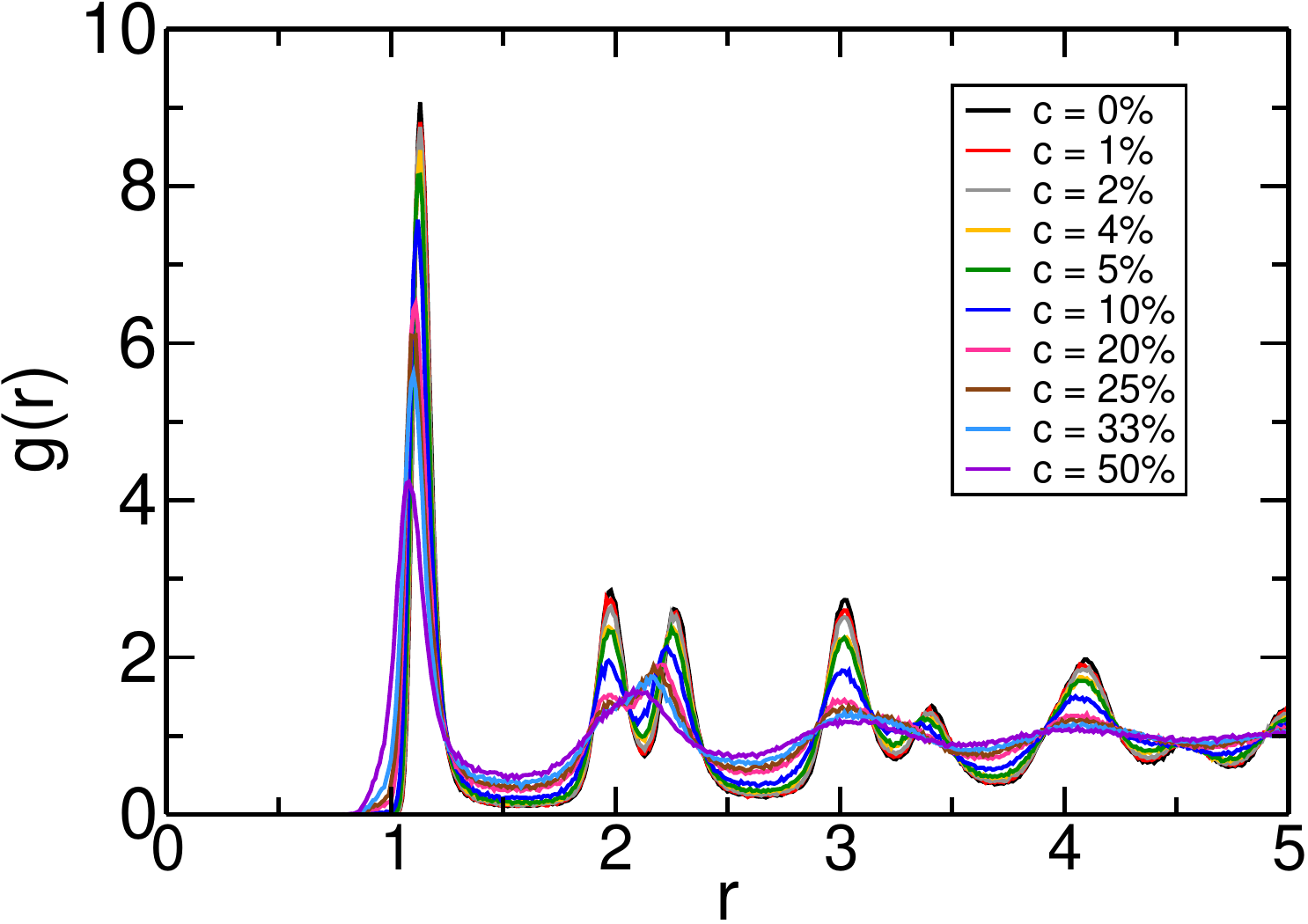}
    \caption{Radial distribution function at $T=0$ for a monodisperse system interacting via LJ$(12,6)$ potential and constituting 0\%, 1\%, 5\%, 10\%, 20\% and 50\% pinned particles. add c\%}
    \label{fig:Ch4-1}
\end{figure}

We examine the RDF for systems containing varying fractions of pinned particles at $T=0$. In the absence of any pinning, the system naturally evolves into an ideal crystalline structure upon cooling to $T = 0$. However, introducing pinned particles disrupts this order, resulting in increasingly disordered configurations as the pinning concentration grows. The RDF plots for systems interacting via the LJ potential are presented in Figure \ref{fig:Ch4-1} for pinning concentrations  $c=0\%, 1\%,2\%,4\%,5\%,10\%,20\%,25\%,33\%, 50\%$. For the unpinned case $(c=0\%)$, the RDF displays a pronounced and narrow peak with the highest amplitude, indicative of a well-ordered crystalline phase. With increasing pinning percentages, the peaks become progressively broader and their heights decrease, reflecting a systematic growth in structural disorder. Furthermore, the position of the first peak of RDF, which corresponds to the minimum of the interparticle potential, is observed to shift towards the left for systems with $c\geq20\%$. At high pinning concentration, the likelihood of finding a particle in minimum energy configuration decreases. Thus high concentration of pinned particles not only enhances the disorder in the system, but also disrupts the equilibrium configuration.

\subsection{HEXATIC-ORDER PARAMETER}
\begin{figure}
    \centering
    \includegraphics[width=1\linewidth]{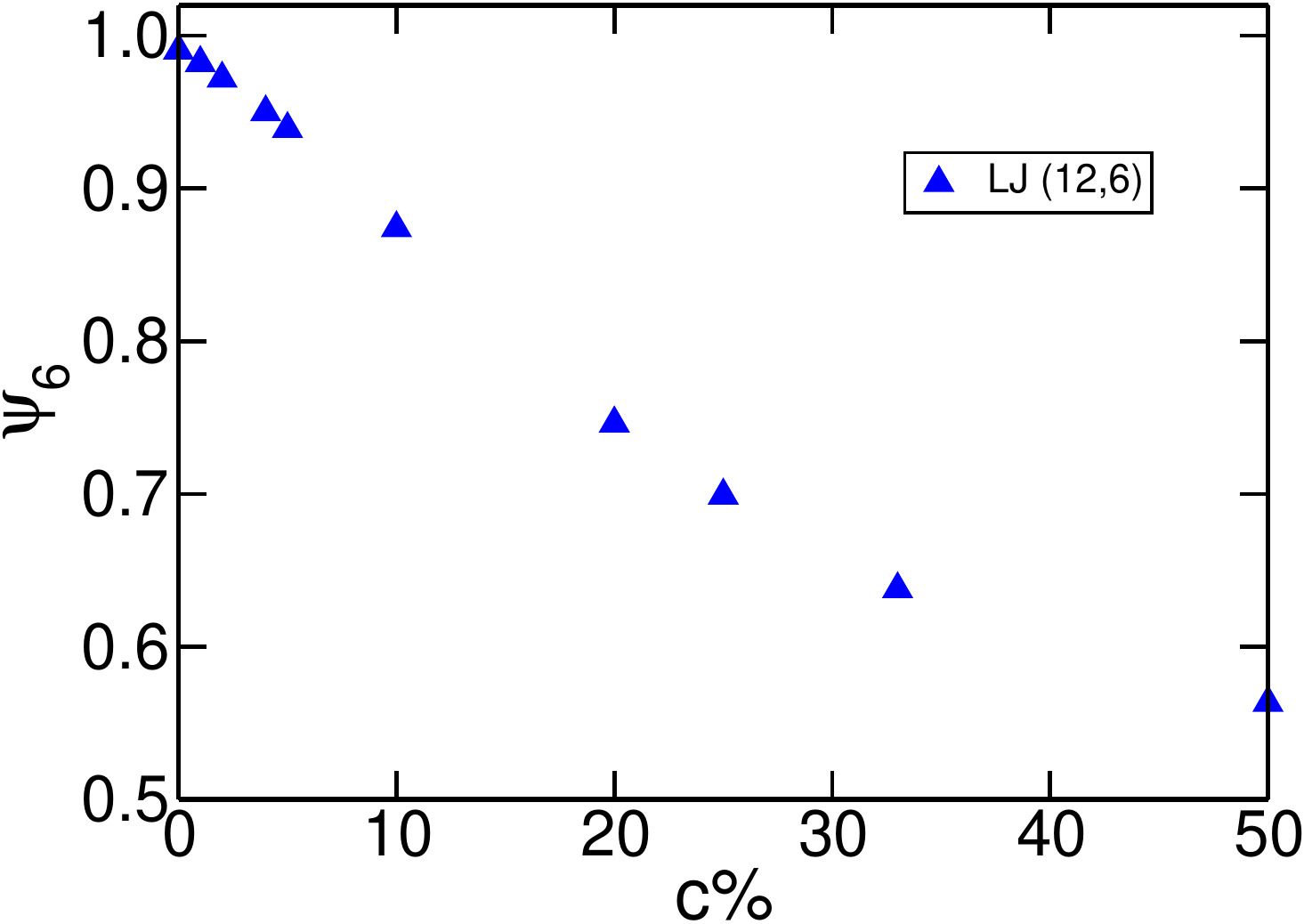}
    \caption{Hexatic-order parameter for an identical particle system with increasing concentration of pinned particles $c\%$. We observe a systematic decrease in orientational ordering with a rise in the concentration of pinned particles.}
    \label{fig:Ch4-2}
\end{figure}
The degree of orientational ordering among the particles in two-dimensional systems is quantified using the global hexatic-order parameter \cite{shuv-surface-structure}. The mathematical expression of the hexatic-order parameter takes the form \cite{halperin1978theory,weber1995melting},
\begin{equation}
    \Psi_6 = \left\langle \frac{1}{N} \left| \sum_{l=1}^{N} \frac{1}{N_b} \sum_{m=1}^{N_b} \exp(i6\theta_{lm}) \right| \right\rangle,
\end{equation}
where $\theta_{lm}$ denotes the angle between the bond connecting particles $l$ and $m$ and a fixed reference axis. Here, $N_b$ represents the number of nearest neighbors associated with each particle, determined by identifying all particles within a cutoff radius. This cutoff is chosen as the distance corresponding to the minimum following the first peak of the radial distribution function.

\begin{figure*}
    \centering
    \includegraphics[width=1\linewidth]{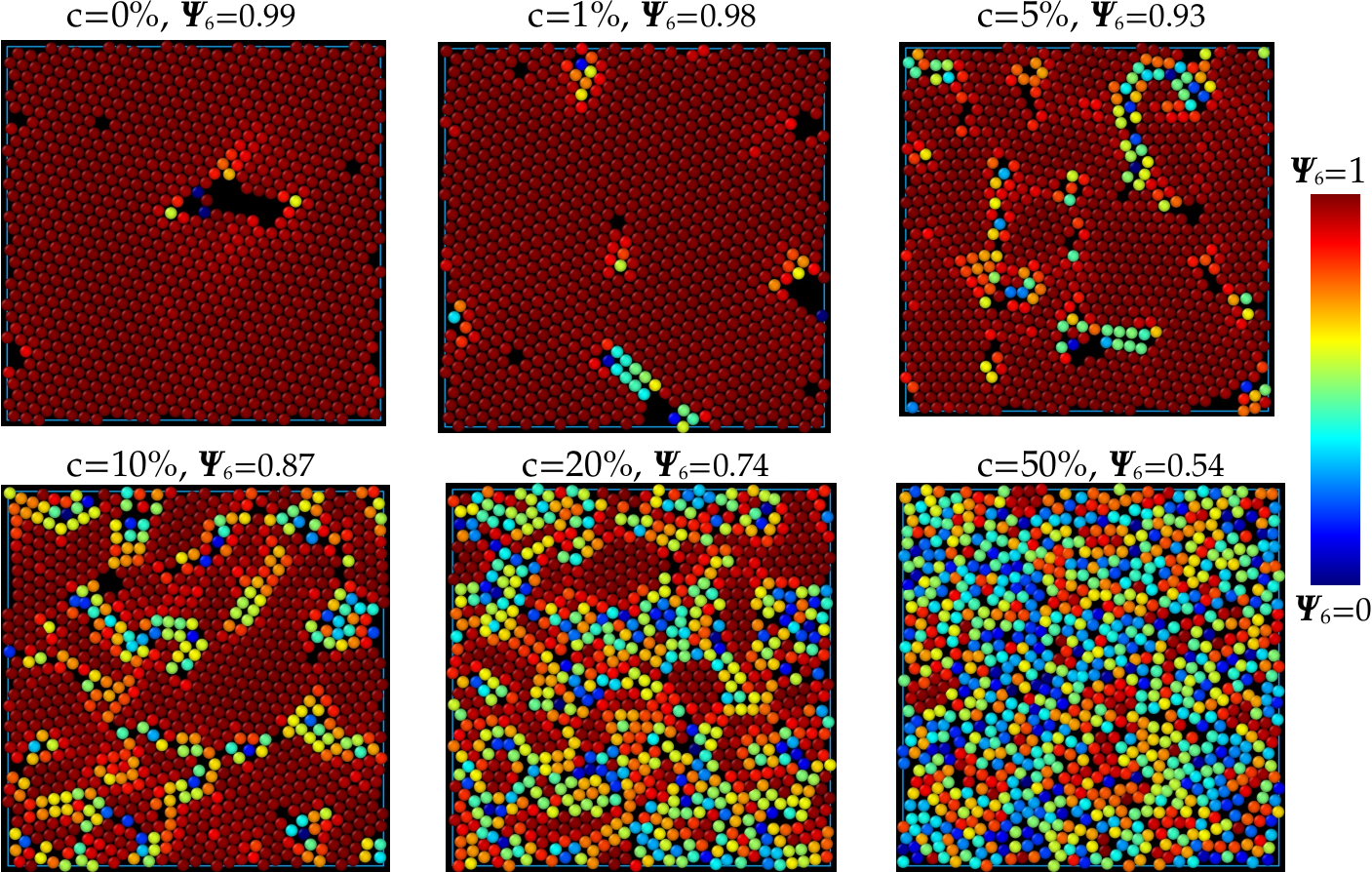}
    \caption{Snapshots showing hexatic-order parameter for one-component system composed of $c=0\%, 1\%, 5\%, 10\%, 20\%$ and $50\%$ pinned particles. The maroon color dots mark the particles surrounded by a group of perfect hexagons of neighboring particles. As we proceed from systems composed of $0\%$ to $50\%$ pinned particles, the number of maroon color particles decreases, indicating the monotonic decrease of the hexatic-order parameter. }
    \label{fig:Ch4-3}
\end{figure*}

Figure \ref{fig:Ch4-3} displays the variation of the hexatic-order parameter across systems with different fractions of pinned particles. In a completely unpinned system, where all particles are identical, the system cools into a crystalline state at $T = 0$, exhibiting a hexatic-order parameter close to unity. However, the inclusion of pinned particles disrupts this long-range order, driving the system toward a hexatic phase, which is an intermediate state exhibiting partial orientational order between crystalline and disordered liquid phases.

As the pinning concentration increases, the value of $\Psi_6$ decreases consistently, indicating a progressive loss of orientational coherence. For a clearer understanding of this behavior, we visualize particle configurations at $T = 0$ in Figure \ref{fig:Ch4-3}. The snapshots for systems with $c=0\%, 1\%, 5\%, 10\%, 20\%$ and $50\%$, each with $\Psi_{6}=0.99, 0.98, 0.93, 0.87, 0.74,0.54$ are shown. In these snapshots, particles forming near-perfect hexagonal arrangements are highlighted in maroon. It is evident that the number of such well-ordered particles declines monotonically with increasing pinning concentration, reflecting the growing disorder introduced by the pinned sites. 

A systematic reduction in both the radial distribution function and the hexatic order parameter is observed with increasing pinning concentration, reflecting a progressive enhancement of structural disorder. This behavior is consistent with experimental observations by Sun et al., who reported a monotonic increase in structural disorder upon increasing the pinning density \cite{Sun_2021}.

\subsection{SELF-INTERMEDIATE SCATTERING FUNCTION}
\begin{figure}
    \centering
    \includegraphics[width=1\linewidth]{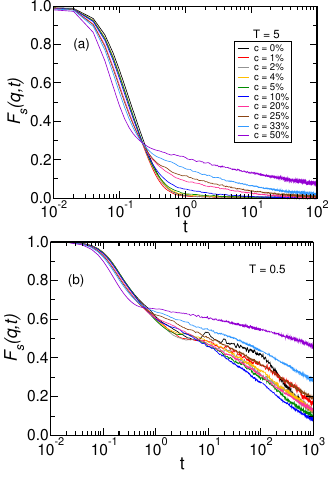}
    \caption{Self-intermediate scattering function for a group of particles interacting with LJ$(12,6)$ potential and consisting of various concentrations of pinned particles (a) at $T = 5$ and (b)  at $T = 0.5$.}
    \label{fig:Ch4-4}
\end{figure}

The self-intermediate scattering function (SISF) serves as a key quantity for probing particle dynamics in Fourier space. It encapsulates combined spatial and temporal information by monitoring how an individual particle evolves over time \cite{Kurzthaler2016}. Specifically, SISF measures the correlation between position of the particle at the initial time and its position at a later time $t$. In a pinned system, the SISF is calculated for the mobile particles as follows, \cite{2015Li,SSuvarna2025POF},
\begin{equation}
    F_{s}(q,t) = \frac{1}{N - N_P} \left\langle \delta\rho({\bf q},t)\, \delta\rho^*({\bf q},0) \right\rangle,
\end{equation}
where $F_s(q,t)$ represents the time-dependent correlation of density fluctuations among the mobile particles. Here, $N$ is the total number of particles in the system, and $N_P$ denotes the number of immobile particles, making $N - N_P$ the count of dynamic particles. The density fluctuation term is defined as $\delta\rho({\bf q},t) = \sum_{j=1}^{N - N_P} \exp[i{\bf q} \cdot {\bf r}_j(t)]$, where ${\bf r}_j(t)$ indicates the position of the $j^{\textrm{th}}$ mobile particle at time $t$, and $q$ is the wave number used for the Fourier transformation. This wave number is given by $q^* = 2\pi / r_1$, with $r_1$ corresponds to the position of the first peak of $g(r)$.

The SISF is computed over a temperature range of $0.4 \leq T \leq 5$ to explore how particle dynamics vary with temperature. The analysis is performed on a well-equilibrated system after cooling it to the desired temperature. Figures \ref{fig:Ch4-4}(a) and \ref{fig:Ch4-4}(b) show the self-intermediate scattering function as a function of time for systems of identical particles interacting via the Lennard–Jones potential, at $T=5$ and $T = 0.5$, respectively. At $T = 5$, the SISF exhibits a simple exponential decay. The decay becomes progressively slower with increasing pinning concentration, indicating a systematic slowdown of the dynamics. At this temperature, around $t=0.3$, we observe a crossover in the decay behavior of SISF with $c\%$. Uptil $t=0.3$, time taken by SISF to decay decreases with rise in pinning concentration. However, beyond this time, the trend reverses. Now, the SISF for systems laiden with more number of frozen particles take longer to decay.

In contrast, at $T = 0.5$, the SISF displays a two-step relaxation behavior, as observed in Figure \ref{fig:Ch4-4}(b). At this temperature, initially, the SISF decays rapidly due to ballistic motion of the particles. This is followed by the emergence of a plateau at intermediate times, reflecting the transient caging of particles by their neighbors \cite{beta-relax1}. At longer times, the system enters the diffusive regime, where particles come out of cages, and exhibit enhanced motion. This is marked by the decay of the SISF \cite{Bongsoo_Kim_92,log_MSD_alpha_beta_Karmakar_2016}. The decay of the SISF shows a non-monotonic dependence on the pinning concentration. As the pinning fraction increases from $c = 4\%$ to $c = 10\%$, the decay becomes faster, indicating enhanced dynamics. However, upon further increasing the pinning concentration to $c = 20\%, 25\%, 33\%$, and $50\%$, the SISF decay progressively slows down, signaling the suppression of particle mobility.

\subsection{RELAXATION MECHANISM}
\begin{figure}
    \centering
    \includegraphics[width=1\linewidth]{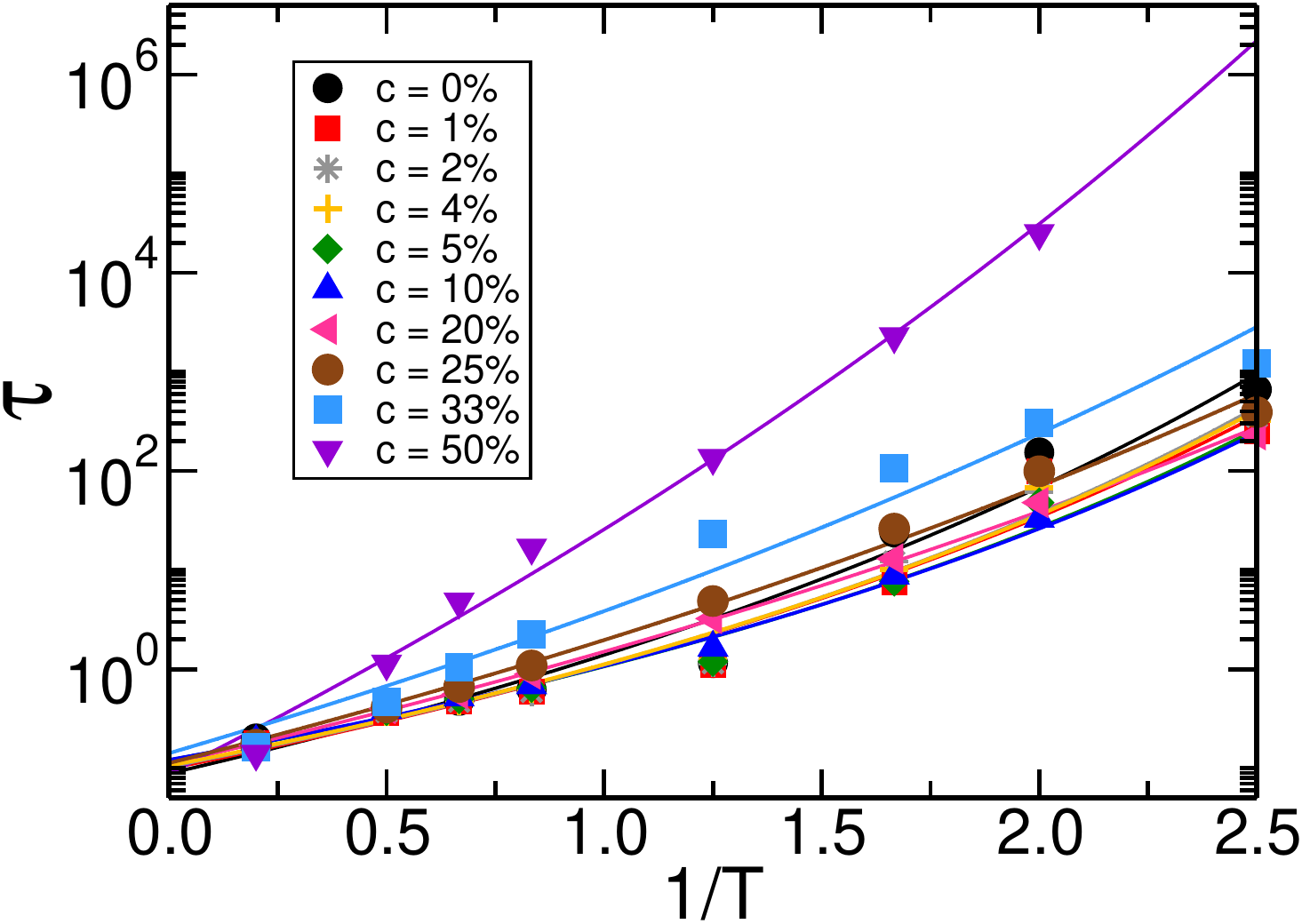}
    \caption{The VFT fit to relaxation time $\tau$ with inverse of temperature for different pinning concentrations. Particles here are subjected to the LJ$(12,6)$ potential. The fitting extracts the value of fragility.}
    \label{fig:Ch4-5}
\end{figure}
During the dynamic evolution of a system, the particle configuration does not remain constant but instead changes continuously due to ongoing particle motion and structural rearrangements. The timescale over which particles reorganize to a state such that the average distance between their nearest neighbors relaxes with time is referred to as the relaxation time. This timescale is mathematically defined as the time at which the SISF decays to $1/e$ of its initial value.

The temperature dependence of the relaxation time provides insight into the kinetic nature of the system and is used to characterize fragility \cite{strong_and_fragile_glass}, a measure of how rapidly the dynamics slow down as the system approaches the glass transition \cite{Sastry1998_Arrhenius}. The fragility index is extracted from the slope of the relaxation time when plotted against the inverse of temperature. To quantify this, we analyze the temperature dependence of relaxation time $\tau$ for a one-component LJ system at various pinning concentrations, as shown in Figure \ref{fig:Ch4-5}. The $\tau \; versus \; 1/T$  data is fitted using the Vogel-Fulcher-Tammann (VFT) relation, expressed as \cite{Chakrabarty_2015},
\begin{equation}
    \tau = \tau_{\infty} \exp\left[\frac{1}{K\left(\frac{T}{T_{\text{VFT}}} - 1\right)}\right],
\end{equation}
where $\tau_{\infty}$ represents the relaxation time in the high-temperature limit, $T_{\text{VFT}}$ is the VFT temperature, and $K$ denotes the fragility parameter.

\begin{figure}
    \centering
    \includegraphics[width=1\linewidth]{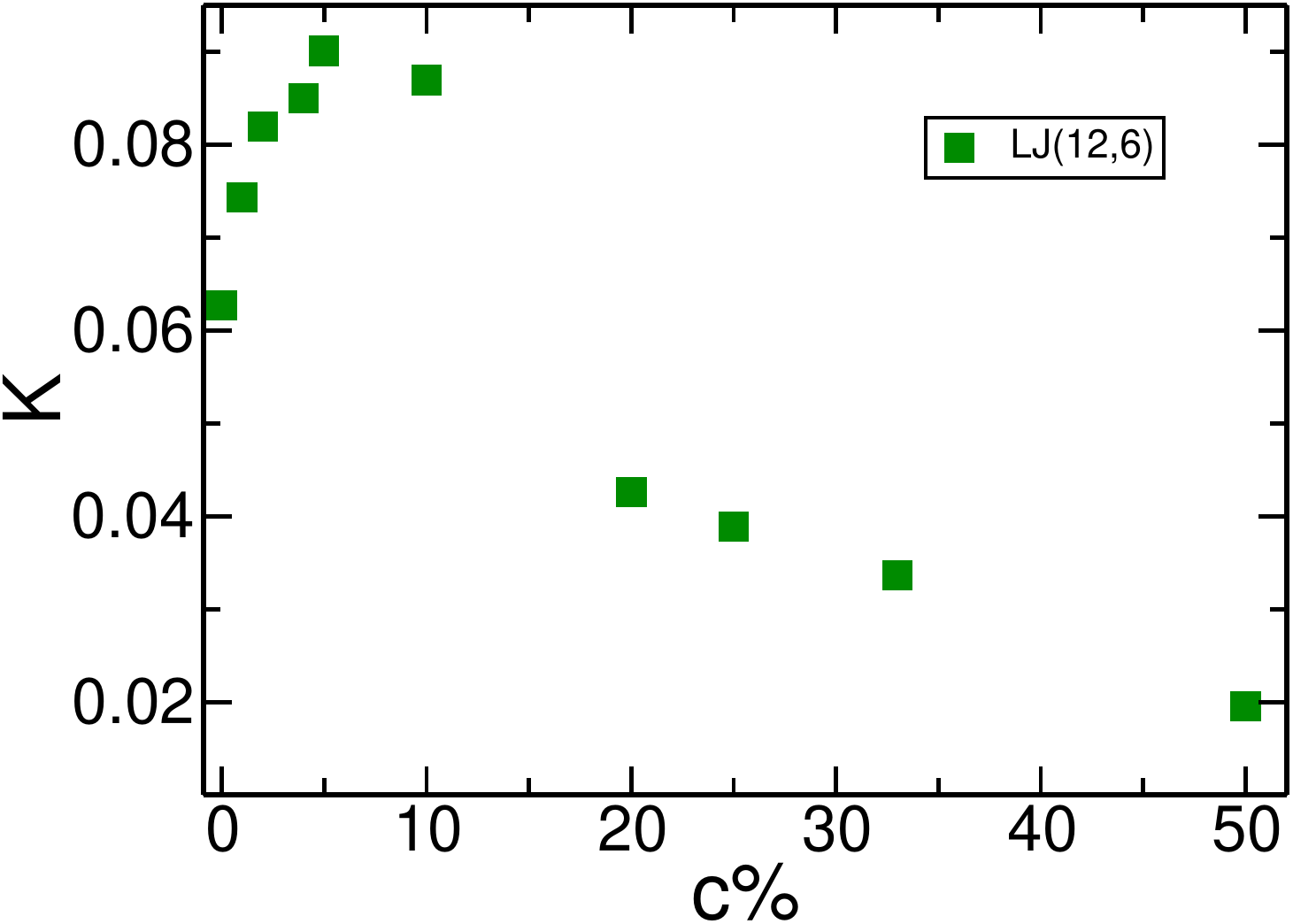}
    \caption{Fragility as a function of percentage of pinned particles, interacting through LJ(12,6) potential. Fragility first rises, attains a maximum value for $c = 10\%$, and then declines with a further rise in $c\%$.}
    \label{fig:Ch4-6}
\end{figure}

Earlier works have demonstrated that increasing the number of pinned particles in a binary glass forming system leads to a reduction in fragility \cite{Chakrabarty_2015,SSuvarna2025POF}. In this work, we explore how fragility changes in a one-component system as the concentration of pinned particles is systematically increased. Figure \ref{fig:Ch4-6} presents the variation of the fragility index $K$ with pinning percentage $c\%$. Interestingly, our results reveal a non-monotonic behavior, where, fragility initially increases as the fraction of pinned particles rises, reaching a maximum around $c = 10\%$. Beyond this point, a further increase in pinning concentration leads to a decline in fragility. This behavior marks a novel observation in one-component systems. 

\subsection{MEAN-SQUARED DISPLACEMENT}
\begin{figure}
    \centering
    \includegraphics[width=1\linewidth]{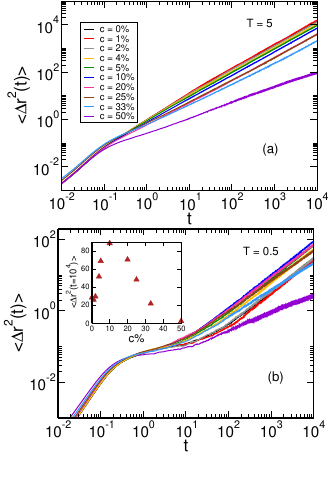}
    \caption{Mean-squared displacement for all the studied pinned systems at temperatures $(a)$ $T = 5$ and $(b)$ $T = 0.5$, \textit{Inset:} Value of mean-squared displacement at $t=10^4$ as a function of pinning concentration.}
    \label{fig:Ch4-7}
\end{figure}

The mean-squared displacement (MSD) of mobile particles in a pinned system is calculated as follows \cite{Bhowmik2016},
\begin{equation}
    \langle \Delta r^{2}(t) \rangle = \frac{1}{N - N_P} \left\langle [ \mathbf{r}_i(t) - \mathbf{r}_i(0) ]^{2} \right\rangle,
\end{equation}
where $N$ is the total number of particles, $N_P$ is the number of pinned particles, and $N - N_P$ thus represents the count of mobile particles. The calculations are performed for time intervals extending up to $t = 10^4$, starting from an equilibrated configuration.


Figures \ref{fig:Ch4-7}(a) and (b) show the MSD curves for $T = 5$ and $T = 0.5$, respectively. At high temperature, particles move larger distances and exhibit enhanced mobility, leading to higher MSD values. In Figure \ref{fig:Ch4-7}(a), MSD at $T=5$ (highest observed temperature) as a function of time $t$ is shown for systems with varying pinning concentration.  Though the overall MSD is high at this temperature, we observe a steady decrease in particle mobility with rising pinning concentration, with the largest displacement occurring for $c = 1\%$, followed by other higher pinning concentration. 

MSD for systems with various $c\%$ at $T=0.5$ is displayed in Figure \ref{fig:Ch4-7}(b). Here, initially MSD $\propto$ $t^{2}$, marking the ballistic motion of the particles. A distinct shoulder in the MSD curve at intermediate times is observed, which marks the localized trapping of particles within cages formed by neighboring particles. With time, the particles come out of the cages and show diffusive motion marked by MSD $\propto$ $t$. We analyze the of MSD obtained at $t=10^4$, to study the dependence of the overall mobility on $c\%$. The plot of values of MSD at $t=10^{4}$ as a function of $c\%$ is shown in the inset of Figure \ref{fig:Ch4-7}(b). $\Delta r^{2}(t=10^{4})$ increases from $c = 4\%$ to $c = 10\%$, reaches a maximum value, and then declines for $c\geq 20\%$. This pattern mirrors the non-monotonicity observed in the fragility index curve shown in Figure \ref{fig:Ch4-6}.




At low temperatures, moderate pinning disrupts local order and facilitates motion, resulting in greater mobility. However, excessive pinning reinforces glassy dynamics. For $c = 50\%$, the plateau of MSD represents an arrested state with negligible long-time motion. This non-monotonic variation in particle mobility demonstrates the delicate balance in two-dimensional identical particle systems, where moderate pinning can relax local constraints and enhance motion, whereas high pinning concentrations dominate the dynamics, driving the system into a glassy state. These observations show qualitative agreement with experimental results reported by Sun et. al. \cite{Sun_2021}.

\subsection{DIFFUSION COEFFICIENT AND ACTIVATION ENERGY}
\begin{figure}
    \centering
    \includegraphics[width=1\linewidth]{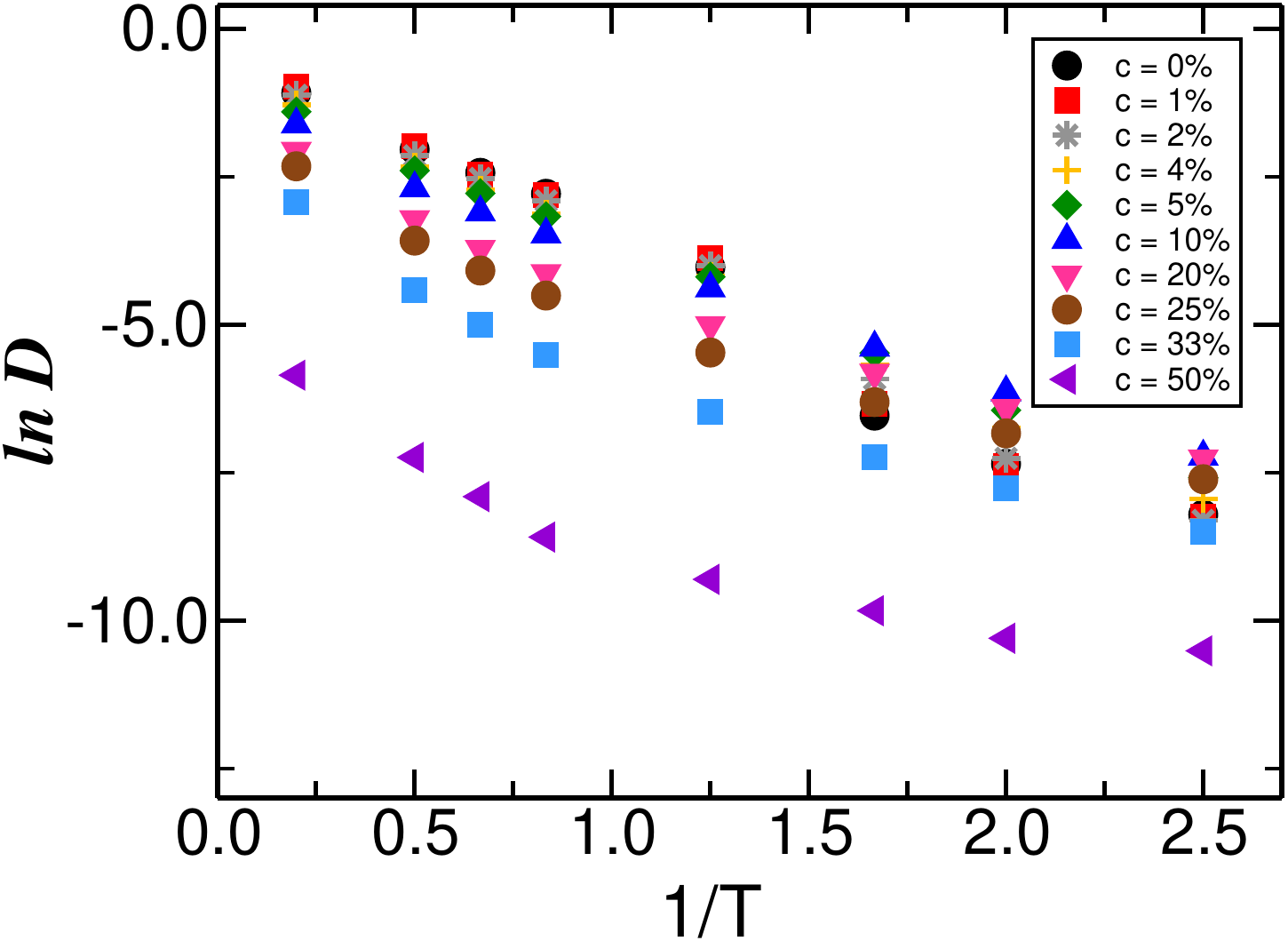}
    \caption{Plot of natural logarithm of diffusion coefficient as a function of inverse of temperature for systems with increasing concentration of pinned particles.}
    \label{fig:Ch4-8}
\end{figure}

The self-diffusion coefficient $D$ is determined from the long-time behavior of the mean-squared displacement. It is mathematically expressed as the long-time slope of MSD, given by \cite{hansen2013theory},
\begin{equation}
    D = \lim_{t \to \infty} \frac{\langle \Delta r^{2}(t) \rangle}{4t}.
\end{equation}
Figure \ref{fig:Ch4-8} presents the variation of $ln D$ with inverse temperature $1/T$ for a one-component system with various pinning fractions. For $c \leq 25\%$, the diffusion coefficient shows a crossover behavior as the temperature decreases from $T = 5$ to $T = 0.4$. In the high-temperature regime (low $1/T$), $D$ is highest for $c = 1\%$ and then decreases systematically. However, this ordering changes as cooling progresses. Around $1/T \approx 1.66$, the curves begin to cross, and by $1/T = 2.5$, the largest diffusion is observed for $c = 20\%$. For the highly pinned systems ($c =33\%,  50\%$), no such crossover is observed. The particle mobility remains low across the entire temperature range for these systems due to strong localization, leading to consistently small diffusion coefficients.

We note that both the diffusion coefficient $D$ and the relaxation time $\tau$ exhibit a data collapse in the high-temperature (low $1/T$) regime when scaled by the pinning concentration $c\%$. This collapse is observed at higher pinning fractions $(c = 20\%, 25\%, 33\%,$ and $50\%$; data not shown), indicating the emergence of a pinning-dominated dynamical regime in which the temperature dependence becomes independent of $c\%$. 


\begin{figure}
    \centering
    \includegraphics[width=1\linewidth]{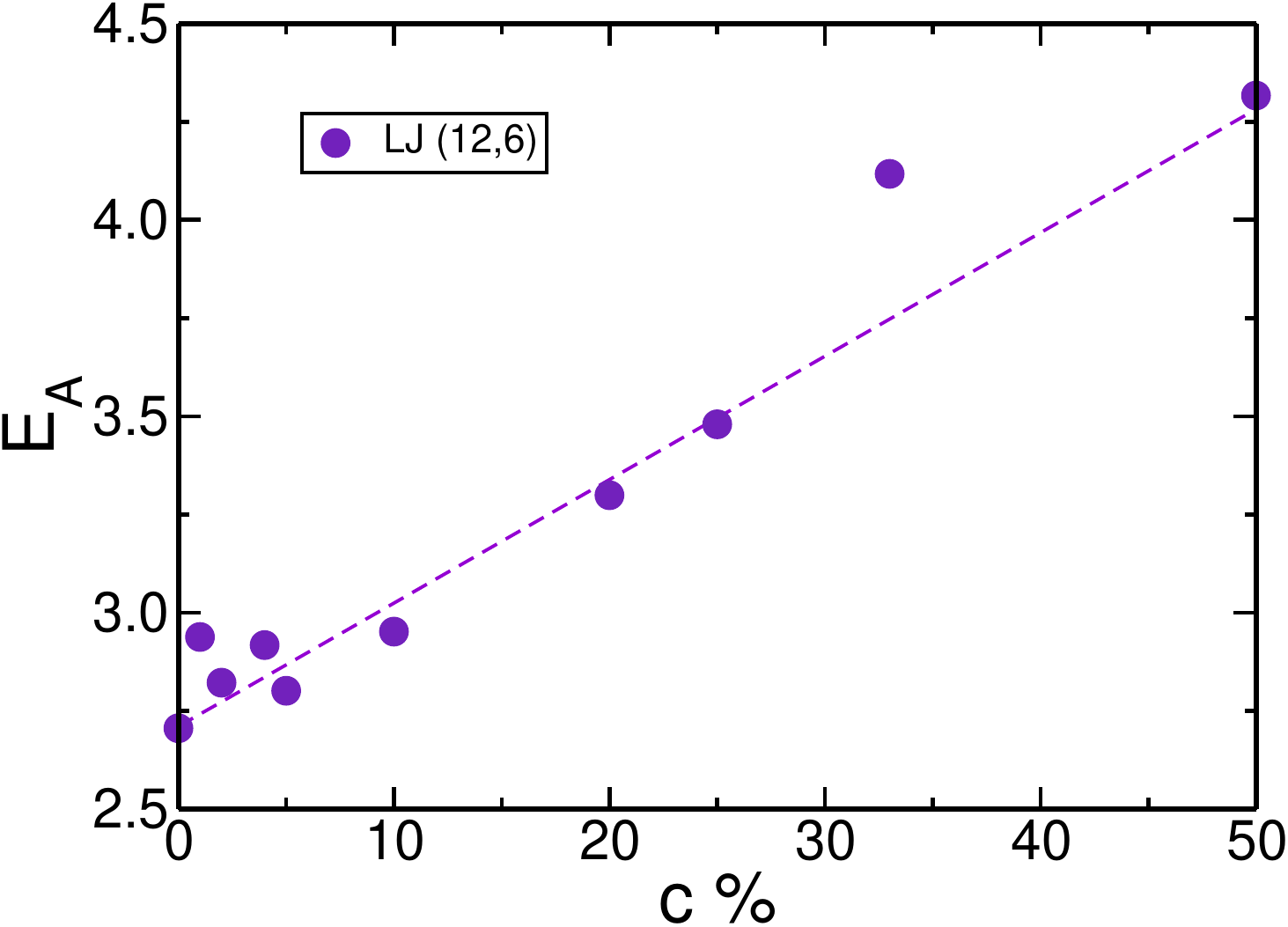}
    \caption{Activation energy $E_{A}$ obtained from the Arrhenius fit to  $ln D$ with $1/T$ for different pinning concentrations.}
    \label{fig:Ch4-9}
\end{figure}

At high temperatures, the self-diffusion coefficient follows an Arrhenius behavior, expressed as,
\begin{equation}
    D = D_{0} \, \exp\left[-\frac{E_A}{k_BT}\right],
\end{equation}
where $D_0$ is the pre-exponential factor corresponding to the maximal diffusivity, $k_B$ is the Boltzmann constant, and $E_A$ denotes the activation energy. Plotting $\ln D$ with $1/T$ yields straight-line behavior in the high-temperature regime, with the slope providing a direct estimate of $E_A$. The extracted activation energies for systems of identical particles with varying pinning fractions are shown in Figure \ref{fig:Ch4-9}. A clear upward trend in $E_A$ emerges as the concentration of pinned particles increases, indicating that immobilization progressively raises the energy barrier for particle motion. However, we observe a fluctuation in the dependence of $E_{A}$ on $c\%$, for $c=1\% \sim 5\%$.

\subsection{NON-GAUSSIAN PARAMETER}
The non-Gaussian parameter ($\alpha_2$) \cite{NGP_First_Paper} is a valuable diagnostic tool for probing dynamical heterogeneity \cite{DH_2000,DH_Book_2011,tanaka2025structural} in particle systems at a fixed temperature \cite{Kim_2011}. In ordinary liquids, displacement and velocity distributions follow the Maxwell–Boltzmann form \cite{hansen2013theory}. In contrast, dynamically heterogeneous systems display departures from Gaussian behavior because some regions contain highly mobile particles while in other regions particles remain trapped for extended periods. This disparity is quantified by $\alpha_2$, which serves as a direct indicator of heterogeneous relaxation dynamics \cite{DynamicHet-1997}. It is defined as,
\begin{equation}
    \alpha_{2}(t) = \frac{d}{d+2} \, \frac{\langle (\Delta r(t))^{4} \rangle}{\langle (\Delta r(t))^{2} \rangle^{2}},
\end{equation}
where $d$ denotes the dimension of the system. It is computed from the ratio of the fourth to the squared second moments of the displacement distribution. A purely Gaussian process yields $\alpha_2 = 0$, whereas positive values signal non-Gaussian behavior, typically associated with transient caging and cooperative motion in glassy and supercooled states \cite{DynamicHet-1997}.

\begin{figure}
    \centering
    \includegraphics[width=1\linewidth]{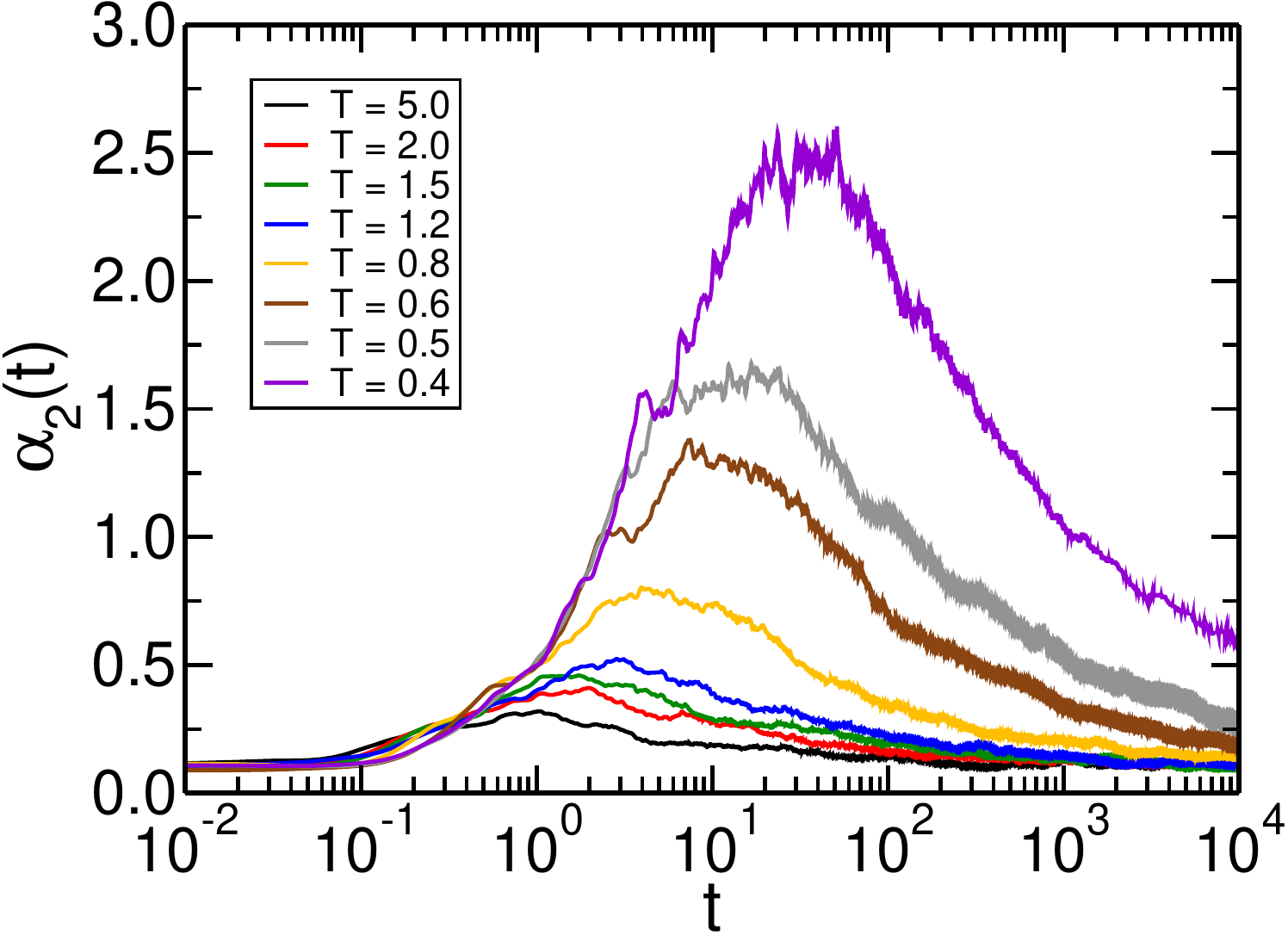}
    \caption{Non-Gaussian parameter with time, obtained at different temperatures for an identical particle system containing $10\%$ pinned particles. The value of $\alpha_{2}$ increases systematically with a decrease in temperature.}
    \label{fig:Ch4-10}
\end{figure}

Figure \ref{fig:Ch4-10} presents the time evolution of $\alpha_2$ for a system with $10\%$ pinned particles at several temperatures. The system is equilibrated sufficiently before tracking the particle dynamics. We observe that the peak height of $\alpha_2$ increases as temperature decreases, reflecting a growth in dynamic heterogeneity upon cooling. This trend indicates that, at lower temperatures, the motion of the system becomes more spatially heterogeneous, with distinct particle subsets exhibiting different mobilities. The rise in peak height of $\alpha_{2}$ with cooling is observed for all the pinning concentrations under study.

At any fixed temperature, the time evolution of the non-Gaussian parameter (NGP) reveals three distinct dynamical regimes experienced by particles. At very short times, $\alpha_2(t) \approx 0$, corresponding to the ballistic motion regime where particles move freely without significant interactions.  As time progresses, particles become transiently trapped within the cages formed by neighboring particles, leading to a rise in the NGP curve. This period coincides with the plateau region observed in the self-intermediate scattering function (SISF) and is referred to as the $\beta$-relaxation regime \cite{Kob1997}. The maximum value of the NGP marks the point of strongest dynamic heterogeneity in the system. Beyond this peak, particles escape their cages through cooperative rearrangements involving correlated motion of many particles. This signals the onset of the $\alpha$-relaxation regime, where the NGP decreases and the dynamics transition to a diffusive nature.

\begin{figure}
    \centering
    \includegraphics[width=1\linewidth]{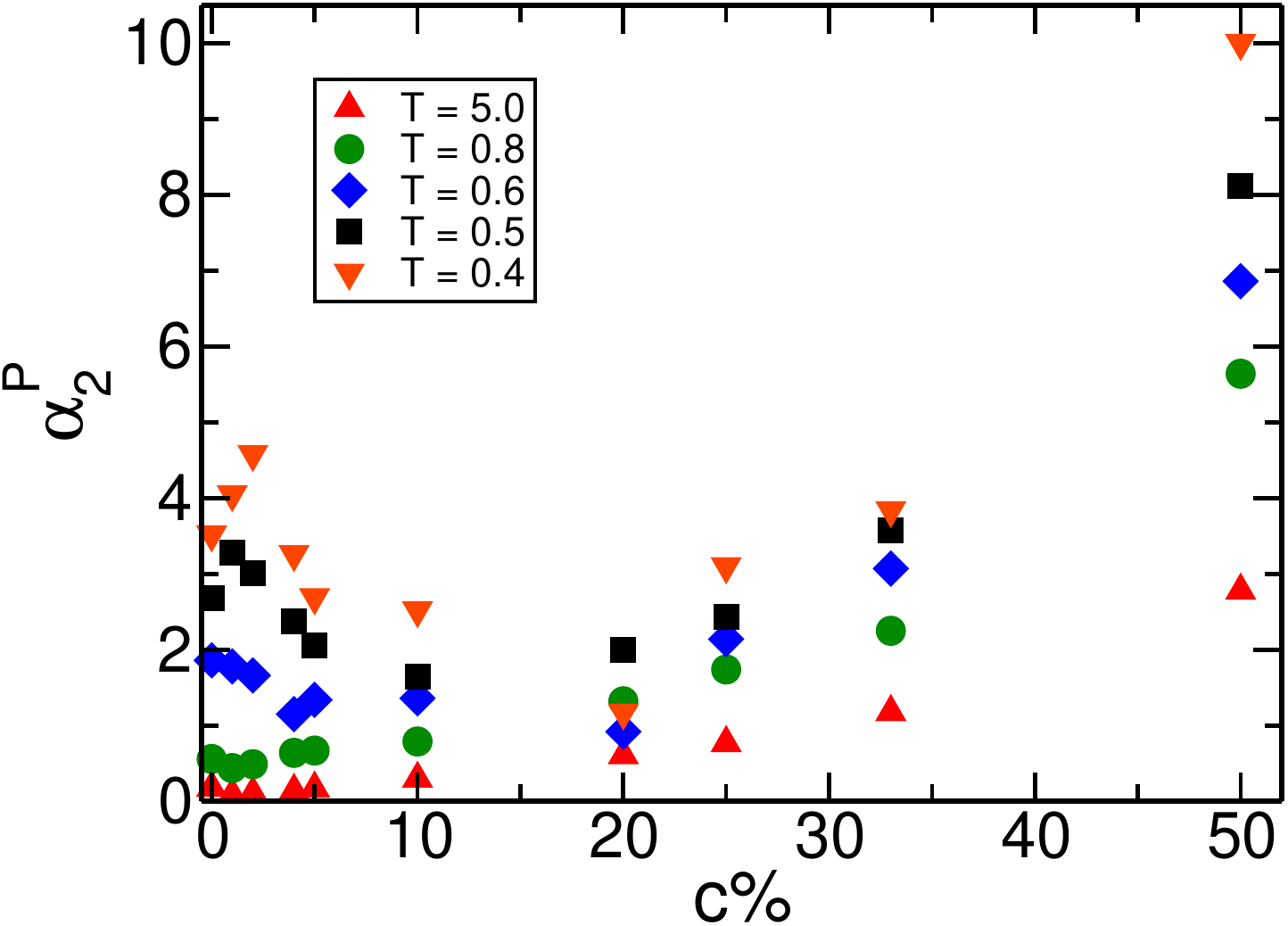}
    \caption{The value of peak height of non-Gaussian parameter at different temperatures $0.4\leq T \leq 5 $, for all the pinned systems under study. At higher temperatures, the peak height is found to be directly dependent on the number of frozen particles. $\alpha_{2}^{P}$ exhibits complex behavior at lower temperatures. }
    \label{fig:Ch4-11}
\end{figure}
In this work, we investigate the variation of dynamic heterogeneity for systems containing different fractions of pinned particles. To capture the overall behavior, we compare the NGP peak height $\alpha_{2}^{P}$ at $T = 5.0, 0.8, 0.6, 0.5, 0.4$ for all the pinning concentrations under study. Figure \ref{fig:Ch4-11} shows how $\alpha_{2}^{P}$ changes with pinning at various temperatures. At high temperatures ($T = 5.0$ and $0.8$), the peak height grows monotonically with increasing $c\%$, indicating that frustration induced by frozen particles enhances dynamic heterogeneity. In contrast, at lower temperatures ($T = 0.6, 0.5, 0.4$), this relationship becomes non-monotonic. Here, the peak height initially decreases when the pinning fraction increases from $c = 4\%$ to $c = 10\%$, reaching a minimum at $c = 10\%$. This suggests that moderate pinning facilitates particle motion and suppresses heterogeneity. Beyond this concentration, further increase in pinning once again leads to higher peak heights, implying a resurgence of dynamic heterogeneity.

Sun et al. probed the dynamic susceptibility $(\chi_{4})$, another key marker of dynamic heterogeneity. In their study, the susceptibility initially increases with pinning concentration and subsequently decreases. Previous works have shown that the susceptibility and the non-Gaussian parameter exhibit opposite trends upon increasing pinning concentration in glass-forming mixtures \cite{2015Li}. In this respect, our results are qualitatively consistent with experimental observations on colloidal systems confined between glass coverslips \cite{Sun_2021}.



\subsection{STOKES-EINSTEIN RELATION WITH TEMPERATURE}
\begin{figure}
    \centering
    \includegraphics[width=1\linewidth]{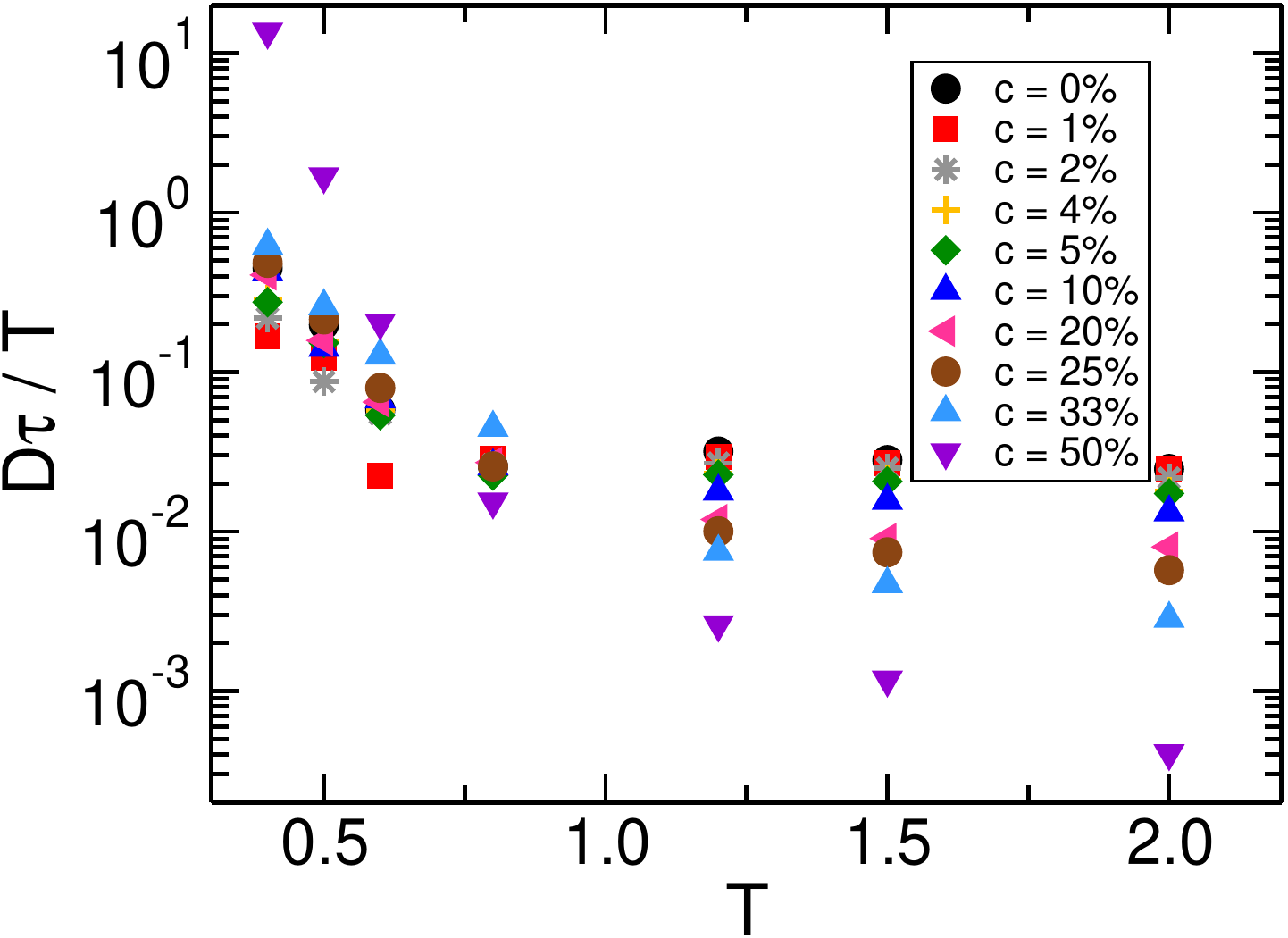}
    \caption{Stokes-Einstein parameter $D\tau/T$ as a function of temperature $T$ shown on a linear-log scale for systems containing $c=0\%,1\%,2\%,4\%,5\%,10\%,20\%,25\%,33\%$ and $50\%$.}
    \label{fig:Ch4-12}
\end{figure}

We examine deviations from the Stokes–Einstein (SE) relation as a function of temperature to probe the emergence of dynamic heterogeneity in the system. The SE parameter, defined as $D\tau/T$, remains constant at high temperatures \cite{SEV_1234,SLS2025_JETP}, consistent with normal liquid behavior. Upon cooling, this parameter increases, signaling a decoupling between the diffusion coefficient $D$ and the structural relaxation time $\tau$. Figure \ref{fig:Ch4-12} presents the temperature dependence of the SE parameter for systems with different pinning fractions, with the SE parameter plotted on a logarithmic scale for clarity, while temperature is shown on a linear scale.

For all pinning concentrations, the SE parameter remains constant above $T \approx 1.2$. Below this temperature, it increases steadily, indicating the onset of heterogeneous dynamics. Notably, for the highly pinned system $(c = 50\%)$, the breakdown of the SE relation occurs at a significantly higher temperature compared to systems with lower pinning fractions. At high temperatures, the SE parameter decreases monotonically with increasing pinning concentration. In contrast, at low temperatures, no simple monotonic dependence on pinning concentration is observed, except at the two extremes, the largest SE parameter is obtained for $c = 50\%$, while the smallest corresponds to $c = 1\%$. A crossover in behavior appears near $T \approx 0.8$, below which the highly pinned system $(c = 50\%)$ exhibits the strongest SE violation, implying the most pronounced dynamic heterogeneity.

\section{CONCLUSIONS}
In this work, we have systematically examined the structural and dynamical properties of a two-dimensional one-component Lennard-Jones liquid subjected to pinning. The introduction of pinned particles induces quenched disorder that significantly alters both equilibrium structure and relaxation dynamics. The rise in the number of pinned particles systematically increases structural disorder within the system, as is evident by the radial distribution function and the hexatic-order parameter. The equilibrium radial distribution function $g(r)$ shows that as the pinning concentration rises, the peaks become broader and their height decreases, indicating a loss of crystalline order. Similarly, the hexatic-order parameter $\Psi_{6}$, which measures the degree of orientational order, systematically decreases with increasing pinning concentration. These results confirm that pinning effectively introduces the frustration necessary to inhibit crystallization. 

The dynamical response of the system, however, exhibits a non-monotonic behavior. At low pinning concentrations $(\sim 4–10\%)$, particle motion is facilitated by the suppression of crystalline order, resulting in faster relaxation and enhanced mobility compared to the unpinned liquid. With further increase in pinning, mobility is progressively reduced, leading to pronounced glass-like dynamics at high pinning fractions $(\sim 20-50\%)$. This is reflected consistently across multiple dynamical measures, including the self-intermediate scattering function and the late-time mean-squared displacement. Importantly, fragility also shows a non-trivial dependence on pinning, increasing at intermediate concentrations before declining at higher ones. The non-monotonic behavior is also captured through the dynamic heterogeneity, as quantified by the non-Gaussian parameter and the Stokes-Einstein parameter. At low temperatures, moderate pinning leads to reduced dynamic heterogeneity by promoting particle motion, whereas higher pinning concentrations lead to a significant increase in heterogeneity due to the restricted movement of mobile particles. 

Taken together, these results demonstrate that pinning serves as an effective mechanism to tune the balance between crystallization and glassy arrest in simple liquids. Our findings are consistent with earlier experimental results on colloidal systems confined between glass coverslips \cite{Sun_2021}, where structural disorder increased with pinning, while the dynamical response exhibited a non-monotonic dependence on the number of pinned particles. Thus confirming that pinning is a versatile and powerful tool for controlling the phase behavior and dynamic properties of two-dimensional liquids. This work not only enhances our fundamental understanding of the glass transition but also has practical implications for designing and controlling material properties in applications ranging from soft matter to advanced materials. 


\section*{ACKNOWLEDGMENT}
P.K.J. acknowledges the financial support from  Anusandhan National Research Foundation (ANRF), India via Grant No. CRG/2022/006365.

\section*{AUTHOR DECLARATIONS}
\subsection*{Conflict of Interest}
The authors declare no conflict of interest.
\subsection*{Author Contributions}
\noindent {\bf Saumya Suvarna}: Formal analysis; Investigation; Methodology (equal); Software; Validation (equal); Visualization; Writing – original draft.
{\bf Prabhat K. Jaiswal}: Conceptualization;  Methodology (equal); Project administration; Supervision; Writing – review \& editing.
{\bf Madhu Priya}:  Conceptualization; Methodology (equal); Project administration; Supervision; Writing – review \& editing. 
\subsection*{Data Availability}
The article contains all the necessary information to reproduce the results presented.

\bibliography{ref}
\end{document}